\documentclass[preprint,12pt]{elsarticle}

\usepackage{amssymb}
\usepackage{graphicx}
\usepackage{epstopdf}
\usepackage{enumerate}

\begin{document}

\begin{frontmatter}



\title{A Xenon Gas Purity Monitor for EXO}

 
\author[umd]{A.~Dobi}
\author[umd]{C.~Hall}
\author[slac]{S.~Herrin}
\author[slac]{A.~Odian}
\author[slac]{C.Y.~Prescott}
\author[slac]{P.C.~Rowson}
\author[slac]{N.~Ackerman}
\author[laurentian]{B.~Aharmin}
\author[bern]{M.~Auger}
\author[stanford]{P.S.~Barbeau}
\author[stanford]{K.~Barry}
\author[csu]{C.~Benitez-Medina}
\author[slac]{M.~Breidenbach}
\author[csu]{S.~Cook}
\author[stanford]{I.~Counts}
\author[umass]{T.~Daniels}
\author[stanford]{R.~DeVoe}

\author[stanford]{M.J.~Dolinski}
\author[laurentian]{K.~Donato}
\author[csu]{W.~Fairbank Jr.}
\author[laurentian]{J.~Farine}
\author[bern]{G.~Giroux}
\author[bern]{R.~Gornea}
\author[carleton]{K.~Graham}
\author[stanford]{G.~Gratta}
\author[stanford]{M.~Green}
\author[carleton]{C.~Hagemann}

\author[csu]{K.~Hall}
\author[laurentian]{D.~Hallman}
\author[carleton]{C.~Hargrove}

\author[itep]{A.~Karelin}
\author[iu]{L.J.~Kaufman}
\author[itep]{A.~Kuchenkov}
\author[umass]{K.~Kumar}
\author[carleton]{J.~Lacey}
\author[seoul]{D.S.~Leonard}
\author[stanford]{F.~LePort}
\author[slac]{D.~Mackay}
\author[ua]{R.~MacLellan}
\author[csu]{B.~Mong}
\author[stanford]{M.~Montero~D\'{i}ez}
\author[stanford]{A.R.~M\"{u}ller}
\author[stanford]{R.~Neilson}
\author[ua]{E.~Niner}

\author[stanford]{K.~O'Sullivan}
\author[ua]{A.~Piepke}
\author[umass]{A.~Pocar}

\author[ua]{K.~Pushkin}
\author[carleton]{E.~Rollin}
\author[sinclair]{D.~Sinclair}
\author[umd]{S.~Slutsky}
\author[itep]{V.~Stekhanov}
\author[stanford]{K.~Twelker}
\author[umd]{N.~Voskanian}
\author[bern]{J.-L.~Vuilleumier}
\author[laurentian]{U.~Wichoski}
\author[slac]{J.~Wodin}
\author[slac]{L.~Yang}
\author[umd]{Y.-R.~Yen}
\address[umd]{Physics Department, University of Maryland, College Park MD, USA}
\address[slac]{SLAC National Accelerator Laboratory, Menlo Park CA, USA}
\address[laurentian]{Physics Department, Laurentian University, Sudbury ON, Canada}
\address[bern]{LHEP, Physikalisches Institut, University of Bern, Bern, Switzerland}
\address[stanford]{Physics Department, Stanford University, Stanford CA, USA}
\address[csu]{Physics Department, Colorado State University, Fort Collins CO, USA}
\address[umass]{Physics Department, University of Massachusetts, Amherst MA, USA}
\address[carleton]{Physics Department, Carleton University, Ottawa, Canada}
\address[itep]{Institute for Theoretical and Experimental Physics, Moscow, Russia}
\address[iu]{Physics Department, Indiana University, Bloomington IN, USA}
\address[seoul]{Physics Department, University of Seoul, Seoul, Korea}
\address[ua]{Dept. of Physics and Astronomy,University of Alabama, Tuscaloosa AL,USA}
\address[sinclair]{Physics Department, Carleton University, Ottawa and TRIUMF, Vancouver, Canada}

\address{}

\begin{abstract}
We discuss the design, operation, and calibration of two versions of a xenon gas purity monitor (GPM) developed for the EXO double beta decay program. The devices are sensitive to concentrations of oxygen well below 1 ppb at an ambient gas pressure of one atmosphere or more.
The theory of operation of the GPM is discussed along with the 
interactions of oxygen and other impurities with the GPM's tungsten filament. Lab tests and experiences in commissioning the EXO-200 double beta decay experiment are described. These devices can also be used on other noble gases.

\end{abstract}


\end{frontmatter}


\section{Introduction}

The EXO collaboration is developing and executing a series of experiments to search for the neutrinoless double beta decay of $^{136}$Xe. The first such experiment, known as EXO-200, is currently taking data at the WIPP facility near Carlsbad, New Mexico, and planning for a next generation experiment is underway. The EXO-200 detector is a liquid xenon TPC, while the successor experiments may be based on liquid or gas technology. One of the primary technical challenges for these detectors is the need to reduce the concentration of electronegative impurities to less than one-part-per-billion, in order to limit the attenuation of charge and light to acceptable levels.

The first step towards controlling the concentration of impurities is to have a simple, robust, and sensitive method to detect those impurities in the gas. We describe here a simple device that we have developed for this purpose which is capable of operating continuously in the EXO xenon gas handling system for the duration of the experiment. We refer to the device as a gas purity monitor (GPM), and it is capable of detecting oxygen at concentrations less than one part-per-billion over a wide range of ambient gas pressures. We have operated GPMs below one atmosphere and above three. It is sensitive to other electronegative impurities at comparable levels. We have installed GPMs at three critical locations in the EXO-200 xenon gas system, and they have proved useful during the commissioning phase of the experiment. The information provided by the GPMs is complementary to other techniques that we use to detect impurities, including occasional mass spectroscopic measurements of the xenon gas, and measurements of the electron lifetime in the liquid xenon. Most importantly, the GPMs provide a highly sensitive and continuous measurement of the xenon gas purity in real time. In this article we describe the design, operation, and calibration of the GPMs. Our other purity monitoring techniques will be described elsewhere.

A tungsten filament immersed in the xenon gas of interest is heated to a high temperature where thermionic emission occurs. A bias voltage placed on the filament allows the emission current to be collected on an anode and measured. In tests performed in our lab with several prototype GPMs, we find that pure xenon gas samples give a relatively large emission current, while impure samples give a negligible current. This demonstrates that the device is sensitive to the impurity concentration in the gas.

At least two physical mechanisms may be responsible for the dependence of the anode current on the gas purity. First, the observed emission current is influenced by space charge build-up around the filament. This current has two components: the primary electron current created directly by the filament and a secondary current due to negatively charged impurity species such as ${\rm O}_2^-$ which form through electron capture reactions. The electron current, typically microamperes in the devices described, dominates.  The ion current, because of the very low velocities of the ions involved, is negligibly small, typically nanoamperes. The ions, however, influence the electron current through space charge effects.

Secondly, impure gas oxidizes the filament, which reduces the emission current by changing the work function of the tungsten surface. Both of these mechanisms lead to high emission currents at high purity, and low emission currents in low purity.

In general, we expect that both mechanisms are at work at the same time, but they can be distinguished from each other based on the GPM behavior that is observed. For example, the signature of the oxidation mechanism is that the emission current decays in a constant-purity gas sample. In this case the time rate of change of the current is the appropriate figure of merit for the purity analysis, while the absolute magnitude of the current indicates the past history of the filament. For the space charge mechanisms, the absolute current is the best measure of the instantaneous purity. Rapid changes of purity, such as can happen in experimental apparatus under use, can be seen in the space charge mode. The oxidation mode can be used to identify species, even those that are not electronegative such as ${\rm N}_2$. The relative importance of the two mechanisms is determined primarily by the filament temperature and bias voltage.

The organization of this paper is the following: We first discuss space charge limited currents in xenon. Then we describe a GPM we constructed in cylindrical geometry and a series of measurements and experiments performed with it. Next we discuss a second GPM, the ``Omega'' style,  constructed for the purpose of studying oxidizing effects on the tungsten filament and its use as a purity monitor.  A series of measurements with the ``Omega'' style GPM is described. Finally, experiences using the cylindrical GPM for commissioning EXO-200 at WIPP are discussed.

\section {Space Charge limited currents in noble gas filled diodes}
\label{sec:SpaceChargeLimitedCurrents} 
\subsection{Space Charge current limits in pure xenon}
Space charge phenomena have been studied for many years in numerous situations involving electric currents, electrostatic potentials, and conducting media including vacuum, gases, liquids, and solids.  In steady state conditions, space charge limited currents can be understood using electrostatic principles and appropriate boundary conditions.

 Consider the case of a vacuum diode consisting of a cylinder of length $L$, radius $b$ with a tungsten filament on its axis of radius $a$. The filament (cathode) is heated by a current to a temperature T, while a voltage +V is applied to the outer wall (anode). Electrons emitted at the surface of the tungsten filament will flow to the anode. The anode current that is collected depends on the temperature T of the filament and the voltage V, as well as the geometry. At low temperature, the current is determined by the thermionic emission from the tungsten. But at high enough temperatures, the current reaches a saturation value set by space charge considerations. This situation was first described by Child and Langmuir and the current-voltage relation is known as the Child's (or Child-Langmuir) Law which predicts a well-known ${\rm V}^{3/2}$ dependence for the space charge limited current.
 
Now consider the case where the  diode is filled with a noble gas such as xenon at pressures of $\sim$ 1 bar. What determines the magnitude of this current?  When the  GPM has been filled with xenon gas, the electron carriers will travel to the anode with a much lower velocity than for the vacuum case. Following the philosophy of Child's Law, we replace the electron velocity with a  value $v_0$ (well measured for xenon gas \cite{biba}).

The space charge limited current can be readily calculated. Starting with $\nabla \cdot {\bf E}= \rho /\epsilon_0$, {\bf j} = $\rho v_{\circ}{\bf e_r}$, and the boundary condition ${\bf E} = 0$ at the cathode, the current is given by
\begin{equation}
\label{eq:SpaceChargeLimitedCurrent}
 J_{electron} = {{2\pi \epsilon_0 v_0 VL} \over {b-a-a\ ln(b/a)}}
\end{equation}
The current is linear in V, so the GPM has an effective resistance given by $ R_{eff} = ({b-a-a\ ln(b/a)) / \ ( 2\pi\epsilon_0v_0L})$. $R_{eff}$ has a value of approximately 2 Mohms for the cylindrical GPM described in this report.
 
\begin{figure}[t]
\centerline{\includegraphics[width=10cm]{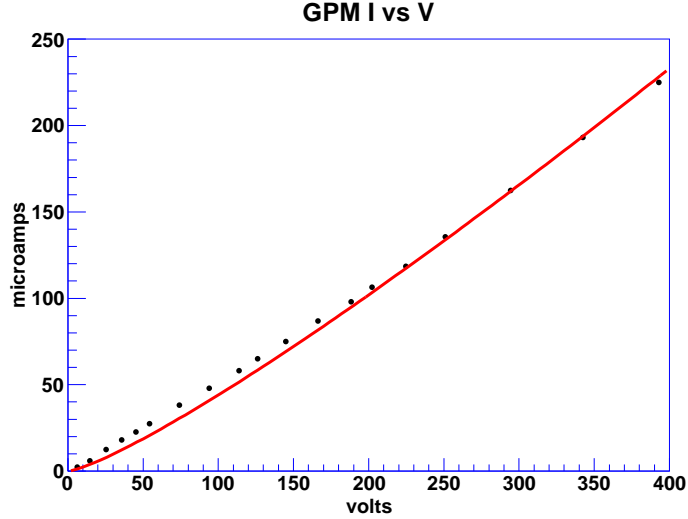}}
\caption{Current versus bias voltage for the cylindrical GPM described in Section \ref{sec:CylindricalGPM}, filled with pure xenon as measured (black points) and predicted for space charge (red curve) using equation (1) and published values of the electron drift velocity.}
\label{figIvsV}
\end{figure}

We have measured the current that flows in the cylindrical GPM described in the next section in this report for values of the applied voltage V, and compared them to the expected current from the formula above.  Figure \ref{figIvsV} shows this comparison.
The data fall  slightly above the predicted values, possibly due to the finite length of the filament and associated end effects.

\subsection{Space Charge currents in the limit of ion current only}
The presence of electronegative impurities, such as ${\rm O}_2$, affect the flow of current by capturing electrons and becoming charged. The drift velocity of the negatively charged impurities is given by $v_{i} = \mu E$, where $\mu$ is the ion mobility. Since typical ion drift velocities are several orders of magnitude below that of electrons, captured electrons contribute a very small amount to the total anode current. In the limit of very impure gas where all electrons are captured near the filament and the current is purely negative ions, the resulting current can be estimated by replacing $v_0$ with $v_{i}= \mu E$ in the space charge equations, giving
\begin{equation}
\label{eq:IonCurrent}
J_{ion}= {2\pi \epsilon_0 \mu V^2L \over (\sqrt{b^2-a^2}-a\ {\rm tan}^{-1}(\sqrt{b^2-a^2}/a))^2}
\end{equation}
which is no longer linear in the voltage, and typically falls in the nanoampere range.

\subsection{Suppression of space charge limited currents by negative ions}
\label{sec:SuppressionByNegIons}
The electron, by being captured on an electronegative impurity, continues to contribute to the space charge, but effectively disappears from the overall current at the anode.  These negative ions remain for a long time (fractions of a second) before reaching the anode.   The ion cloud around the filament builds up because of the low clearing velocity, hence long clearing times. The electron current is suppressed. As the electron velocity is almost constant as a function of the electric field, by lowering the applied voltage V, the effect of the ion cloud can be increased.  The sensitivity of the GPM to small amounts of impurities  stems from the low velocities of the ions, thus ``amplifying'' the ion concentration.

It is easy to observe effects of negative ions on the anode current.  The filament current is initially off. When it is turned on, the anode current has a dramatic time dependence depending on the amount of impurities in the gas.

\begin{figure}[t]
\centerline{\includegraphics[height=8cm]{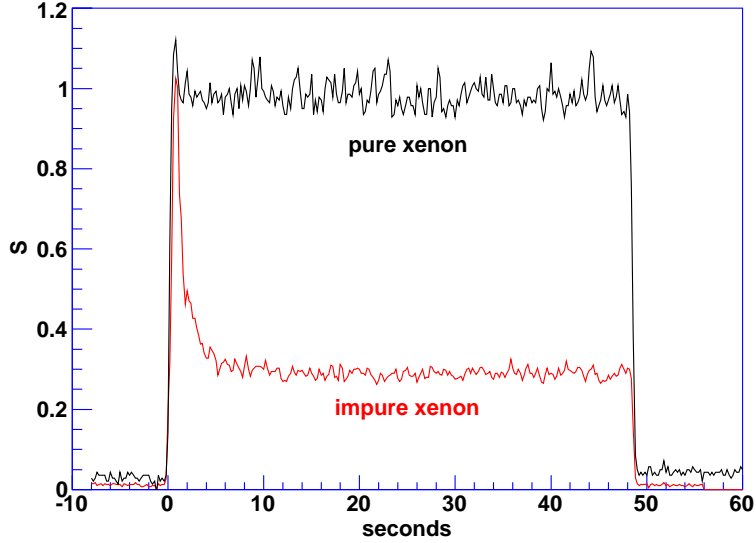}}
\caption{Scope traces of the GPM anode current, plotted as  $S$, the ratio of currents normalized to those for pure xenon,
when the filament is turned on for 50 seconds. The first (in black) shows pure xenon; the second (in red) shows the response initially equal to pure xenon, but then  suppression as an ion cloud builds up in the  xenon containing electronegative impurities. Variations in operation conditions render the precision of pure-xenon normalization to about 10\%. }
\label{figScopes}
\end{figure}

Figure \ref{figScopes} shows scope traces of the anode current for two cases, pure xenon and xenon with impurities added, as the filament is turned on for 50 seconds. For pure xenon the current builds up in a fraction of a second to a constant value, and remains there until the filament current goes away. For impure xenon, the current rises initially to the value for pure xenon, then decreases rapidly in the next few seconds. The ion cloud grows, and the current declines as the ion cloud influences the current through its space charge. Figure \ref{figScopes} shows scope traces of the anode current for these two cases. For impure xenon the scope trace has a leading spike when the filament is turned on, a clear sign of the presence of impurities. Steady state conditions settle in after a few seconds. The scope serves as a simple real-time monitor for impurities, simply by looking at scope traces of the anode current as the filament is turned on.

Relating the measured anode current to the amount of impurity in the xenon gas is a primary goal of interest. Surely they are not linearly related, but what is the relationship? In the Appendix we discuss the space charge limited currents in the case where we have electronegative impurities present and suppressed anode currents.  We show that, under reasonable assumptions, the  density of the {\it neutral} impurity component, $M^{\circ}$, is related to the normalized GPM anode current $S$ by
\begin{equation}
M^{\circ} =k {1-S \over S}
\end{equation}
where $S = J/J_0$  and $J_0$ is the current for pure xenon, and
\begin{equation}
k \approx { 2 e v_{\circ} \mu V  \over \alpha  b^2}
\end{equation}
where $e$ is the charge of the electron, $\mu$ the mobility of the negative ion, and $\alpha$ is related to the probability for an electron attaching to a $M^{\circ}$ molecule, converting to an $M^-$ ion.
 
Since $\alpha$ is an unknown quantity, to assign an absolute value to $M^{\circ}$, the scale needs to be fixed by calibration. Calibration of the GPM would consist of one or more measurements of a known impure sample. Measurement of $S$ on a  sample of known concentration $M^{\circ}$ sets the scale for all values of $S$. This method was used for calibration with a sample of  ${\rm O}_2$ as described in Section \ref{sec:Calibration}. Of course, this calibration would apply to this one species of impurity, and only when other global parameters such as pressure, temperature, bias voltage, etc. are held fixed.

The important information here is the $(1-S)/S$ relationship between the measured $S$ and the impurity concentration $M^{\circ}$. This relation has simple interpretations.  As the electron current is suppressed (and therefore its space charge is suppressed) by a factor of $S$, the ion space charge, $M^-$, replaces the missing electrons by a factor $(1-S)$, and $M^{\circ}$ is proportional to $M^-$.  The $1/S$ term arises because the rate of production of $M^-$ is proportional to the electron current. The constant $k$ has the value of $M^{\circ}$ for which $S$ = 0.5.

Using the relation $v_{i} = \mu V / b$, the coefficient $k$ can be written \[k \approx { 2ev_{\circ}v_{i} \over \alpha  b}.\]
 
Low ion velocity, for a given  normalized GPM current $S$, leads to a smaller $M^{\circ}$ value. This shows why it is important to use low bias voltages, giving low ion velocities, to reach high sensitivity to $M^{\circ}$.  Conversely, low probability coefficients $\alpha$ lead to insensitivity to $M^{\circ}$.
 
 
\section{ The Cylindrical GPM Operation in Space Charge Limited Mode}
\label{sec:CylindricalGPM}
\begin{figure}
\centerline{\includegraphics[width=4cm]{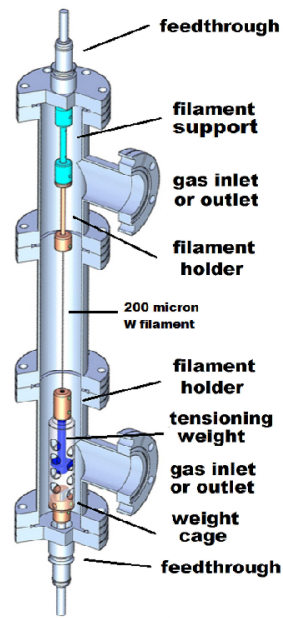}}
\caption{ A cylindrical GPM consisting of 2-3/4 inch conflat vacuum components and a 13 cm long tungsten filament centered by posts and held by set screws. The filament is held under constant tension by a 40 gm weight at the lower end free to slide vertically. This design requires the GPM to be mounted vertically as shown. This is the design used to study space charge limited currents in this report.  Three units are in operation as monitors for EXO-200 at WIPP.}
\label{figCyl}
\end{figure}

The cylindrical GPM is shown in Figure \ref{figCyl}. The filament in the cylindrical GPM is a 200 micron diameter tungsten wire, and its length is 13 cm. The body is constructed of  vacuum components available from a catalog, while the internal parts are machined in our shops. The xenon gas flows through the device and across the filament during operations.

\begin{figure}
\centerline{\includegraphics[width=11cm]{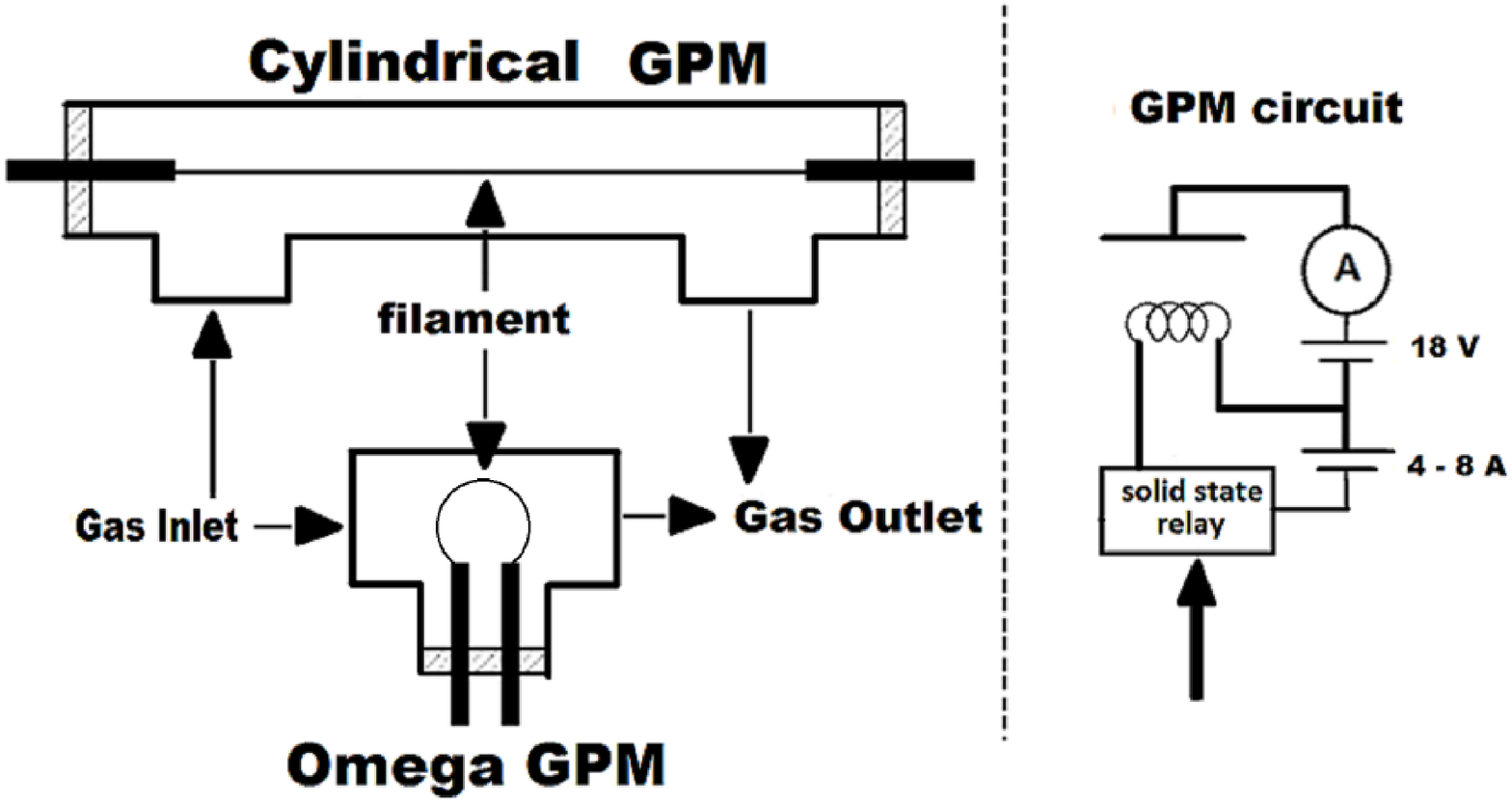}}
\caption{Left: diagram of the two GPM geometries (Cylindrical and Omega). Right: simplified GPM electrical circuit. The filament is heated by applying 10 volts for 4 amperes in the case of the Cylindrical GPM for operating in the space charge limited mode, or 2 to 5 volts for 4-8 amperes (depending on the filament resistance and the desired temperature) for the Omega GPM. The anode current flowing due to a bias voltage of -18 V is measured. The stainless steel plumbing is used as the collection anode.  A computer drives the solid state relay to turn the filament current ON.}
\label{figSchem}
\end{figure}

We heat the filament by applying a voltage, regulated by current, across the two feedthroughs. The temperature of the filament can be determined from the resistivity from published tables or by using the following empirical formula: $[T/T_0]^{1.19} = [R/R_0]$, where $T$ is the final temperature, $T_0$ is room temperature and $R$ and $R_0$ the resistances, both measured in degrees Kelvin. The 13 cm long filament in the cylindrical GPM operates in these studies at 4.0 A at 10.5 V and a temperature of 2150 K. A simplified schematic is shown in Figure \ref{figSchem}.

The ON/OFF status of the filament current is controlled by a computer.  It is normally OFF, and cycled ON for a period of 15 seconds (or longer if desired) and the GPM anode current sampled during this period after waiting for $\geq 5$ seconds to allow the anode currents to settle. The cycles can be repeated as often or as infrequently as desired.

\subsection{Tests on electronegative impurities}
\begin{figure}
\centerline{\includegraphics[height=6cm]{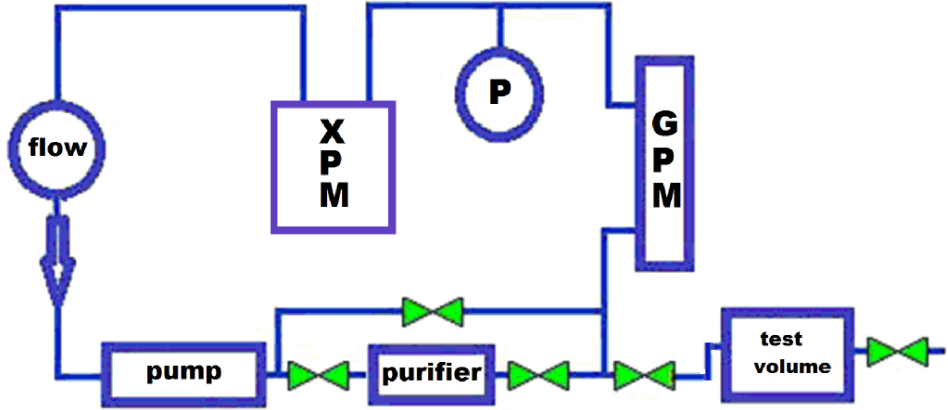}}
\caption{A simplified diagram of the lab test system consisting of a GPM, a liquid xenon purity monitor (XPM), a pump, a purifier, a flow meter, and a pressure gauge.  This system was built as a prototype for EXO-200.}
\label{figSimpLab}
\end{figure}

The GPM was installed in the laboratory in a closed loop xenon system that included a circulating pump, a purifier (which is a heated zirconium getter\footnote{SAES Pure Gas Inc. Model PS3-MT3-R}), a liquid lifetime monitor, a pressure gauge, a flow meter, and test volume on a valved side port that allowed for injection of known impurities into the gas stream. A simplified diagram of the lab system is shown in Figure \ref{figSimpLab}.  Not shown in the diagram, but important to the use, are the supply bottles, vacuum  pumps, and a cryogenic system.

\begin{figure}
\centerline{\includegraphics[height=8cm]{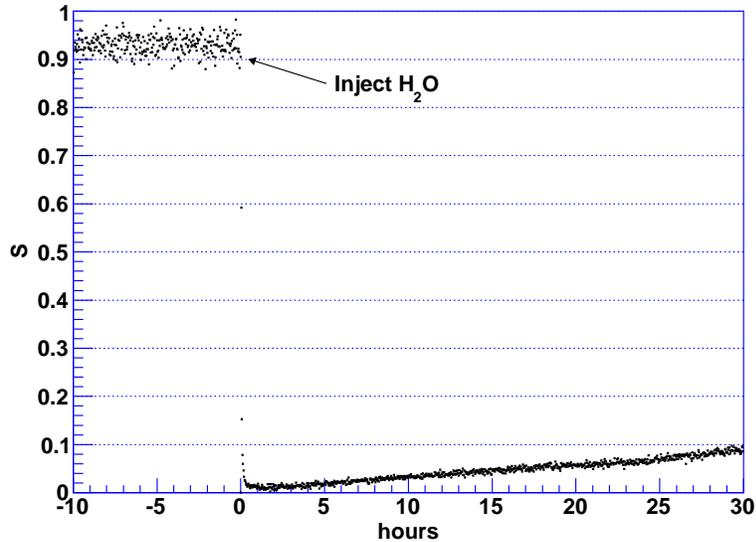}}
\caption{GPM response to injection of water vapor into the test system}
\label{figWater}
\end{figure}

Initial tests were done with several different gas samples, including ${\rm O}_2$, ${\rm H}_2{\rm O}$ vapor, air, and HFE7000\footnote{HFE7000 (formula ${\rm CF}_3{\rm CF}_2{\rm CF}_2{\rm OCH}_3$) is a heat transfer fluid manufactured by the 3M company and used in the EXO-200 cryostat.}. Figure \ref{figWater} shows the response to  water vapor, at a pressure of 20 $\mu$m Hg from the test volume. At $t=0$, a valve was opened and the water vapor  was allowed to mix with the room temperature xenon as it flowed past the test volume.  The GPM immediately dropped to very low current, a typical response. The current remained suppressed for many hours of circulation. Other gas samples showed similar effects on the GPM current. In the next sections we discuss correlations with electron lifetimes in liquid xenon and calibration of the GPM using ${\rm O}_2$, a known electronegative substance, because it is a gas, unlike water vapor, at liquid xenon temperatures.

\subsection{ Correlation with liquid xenon lifetimes}
The underlying motivation for monitoring the xenon purity in the gas phase is to assist in the purifying of the EXO-200 TPC's liquid xenon during commissioning and data collection operations.  The EXO-200 system at WIPP pumps the xenon around a loop, going from the liquid phase in the TPC, through a heater and into the gas phase, through the pump, then through a purifier before re-liquefying and returning to the TPC.  The GPMs, three in all, monitor the gas purity at three points in this loop (see Figure \ref{figSimpEXO}).

The test setup at our lab was built as a prototype of the WIPP setup, so has many features in common. We have used it to study the correlation of the GPM signal with the electron drift lifetimes in the xenon liquid in this closed loop test system in the lab. The liquid xenon purity  monitor (XPM) is patterned after the ICARUS design \cite{bibb}.

\begin{figure}
\centerline{\includegraphics[height=8cm]{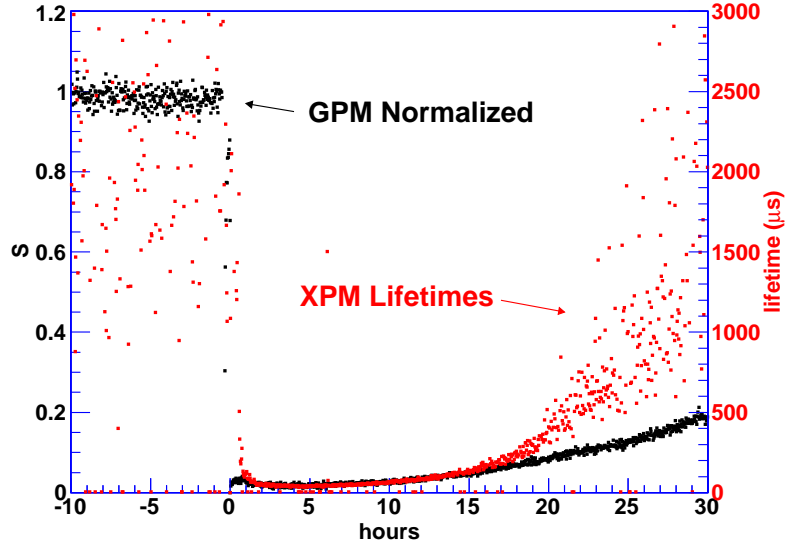}}
\caption{Correlation plot of the  GPM signal (S) and the electron lifetime in liquid xenon upon injection of ${\rm O}_2$}
\label{figCorPlot}
\end{figure}

The test consisted of injections of ${\rm O}_2$ at a pressure of 20 $\mu$m Hg, as measured by a thermocouple gauge, from a test volume of $\sim 140$ ${\rm cm}^3$ into the gas phase. The xenon was then allowed to circulate with the pump on, but with the purifier bypassed, at a rate of $\sim$ 1.5 SLPM (standard liters per minute). Mixing was allowed to occur over the next several hours. The rate of flow around the system was such that a full volume change took about 3 hours. During this period, we monitored the GPM current and the electron lifetimes. The fall in the GPM current, initially rapid, stabilized after several hours, corresponding to full mixing of the ${\rm O}_2$ with the xenon.   Both the GPM response and the electron lifetime responses show a minimum around 5 hours into the test, and clearly are rising by 10 hours. The only hot element in this system is the filament, pulsed ON for 25 seconds to sample the GPM current every 5 minutes.  After 24.3 hours, the bypass at the purifier was closed, and the inline valves at the input and output of the purifier opened, so that the flow passed through the purifier. Purity and lifetimes begin to improve even while the purifier is bypassed, but accelerate when it is in the loop. Figure \ref{figCorPlot} shows the GPM signal and xenon lifetimes, overlayed, for 30 hours of this test.

\subsection{Calibration}
\label{sec:Calibration}
\begin{figure}
\centerline{\includegraphics[height=8cm]{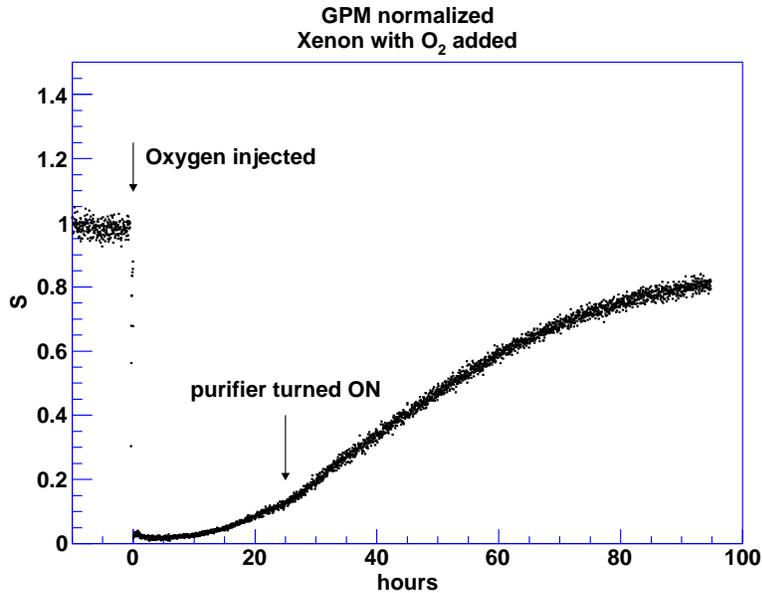}}
\caption{ GPM normalized response,\ $S$, during the injection and subsequent purification of oxygen}
\label{figInjO2}
\end{figure}

As an example of how to calibrate the GPM, we have used the same data as shown in Figure \ref{figCorPlot}. For the calibration the electron lifetime monitor was full of liquid xenon,  approximately 0.6 liters. Initially to clean the xenon, the  system had circulated the gas for a considerable time with the purifier ON, until the GPM read full current ($S = 1$). Oxygen was injected at $t=0$. The fall in the GPM current, initially rapid, stabilized after several hours at a value of $\sim 0.018$, corresponding to full mixing of the ${\rm O}_2$ with the xenon. After a period of $\sim 10$ hours, the GPM current began a slow rise.  At 24.3 hours, the purifier bypass was closed and the purifier became operative. The GPM current continued to rise until the test was terminated at 95 hours, reaching a normalized GPM reading of $S = 0.8$. The normalized  GPM response during this test is shown in Figure \ref{figInjO2} (the same as shown in Figure \ref{figCorPlot}, but extending out to 100 hours).

For calibration purposes, we know the amount of ${\rm O}_2$ injected into the system.  It was 11 $\pm$ 1 ppb if uniformly mixed with xenon. As discussed in Section \ref{sec:OmegaGPM} of this report, ${\rm O}_2$ can be adsorbed by the walls and the tungsten filament quickly, so the amount in the xenon can be somewhat lower.  The lifetime measurement in the liquid xenon can be converted to a concentration using measured rate constants for ${\rm O}_2$ in liquid xenon \cite {bibc}.  At 5 hours into this run, the lifetime measured was 40 $\mu$sec, corresponding to 7.5 $\pm$1 ppb of ${\rm O}_2$.

\begin{figure}
\centerline{\includegraphics[height=8cm]{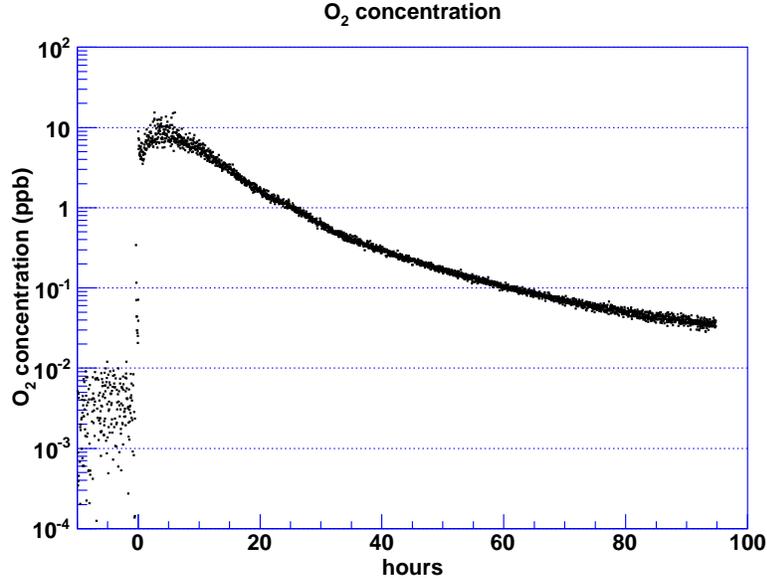}}
\caption{ Estimated ${\rm O}_2$ concentrations based on the GPM calibration and model described in the text and the Appendix }
\label{figO2Conc}
\end{figure}

Assuming the ${\rm O}_2$ was fully and uniformly mixed, the ${\rm O}_2$ concentration  of 7.5 ppb ${\rm O}_2$ gave a GPM reading of 0.018 at 5 hours.  Using the relation $M^{\circ} = k(1-S)/S$ we find that $k = 1.4 \times 10^{-10}$. At the end of the test with the GPM at 0.8, the ${\rm O}_2$ concentration was estimated to be $M^{\circ} = .35 \times 10^{-10}$ (35 parts per trillion). Figure \ref{figO2Conc} shows the estimated ${\rm O}_2$ concentration based on this calibration technique and the GPM model described in the Appendix,  for the duration of this test.
 
In general, calibration may not be easy.  Residual species  in our xenon carrier gas, such as hydrogen, nitrogen, oxygen, carbon monoxide, carbon dioxide, nitric oxide and so on, have varying tendencies to adsorb onto surfaces in the apparatus, affinities that also depend on the precise material and surface quality of the surfaces available. In particular, nonvolatile impurities, most notably water, will not mix well.  They prefer to attach to surfaces or freeze out, so calibration of a nonvolatile impurity by injection of a known amount into the gas stream, using the technique described here, would likely result in an overestimate of the concentration in the gas. Monitoring and purifying  water vapor in a cryogenic system such as EXO-200 can be a difficult problem.
 
\subsection{Dependence on Pressure}
\begin{figure}[t]
\centerline{\includegraphics[height=8cm]{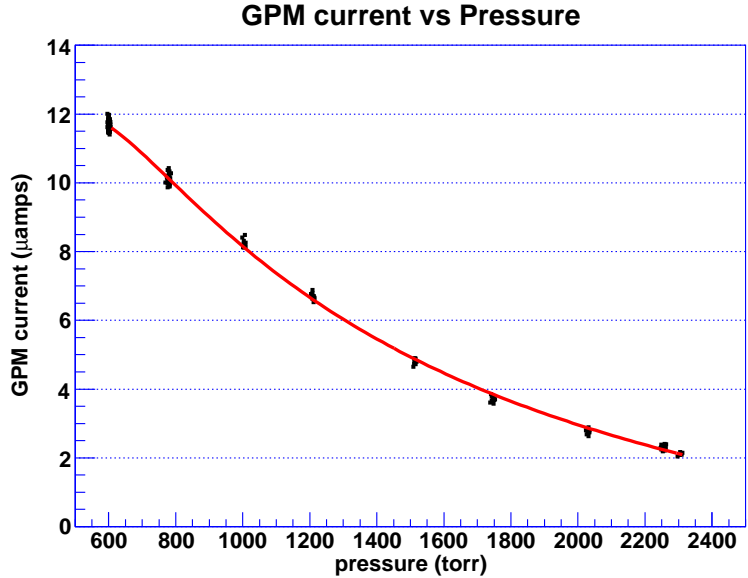}}
\caption{Cylindrical GPM current versus pressure for pure xenon at a bias voltage of -18 V. The curve is a simple fit discussed in the text.}
\label{figPress}
\end{figure}

The linear relationship between GPM current and bias voltage V discussed in Section \ref{sec:SpaceChargeLimitedCurrents} is based on assuming a constant electron drift velocity $v_{\circ}$.  The data, as seen in Figure \ref{figIvsV}, show some deviation from linearity. The non-linearity arises  primarily due to non-constant $v_{\circ}$ as V changes.   The velocity  is  determined well by experiment and is expected to scale with $E/P$\cite{biba}. Thus we would expect the GPM current would depend both on the bias voltage V and the gas pressure $P$.  This dependence is easily measured by varying the system pressure, recording the GPM current, and fitting the data to a curve.  Figure \ref{figPress} shows the resulting data and fit over a pressure range of 600 to 2300 torr.

A simple power law fits the data well.  For the cylindrical GPM in this report, we fit with $J_{fit} = \alpha_1 + \alpha_2/P + \alpha_3/P^2$ , $P$ in torr, and find $\alpha_1 = -4.4279, \alpha_2 = 16982$, and $\alpha_3 = -4.4003\times 10^6$ for the pressure range 600-2300 torr for a bias voltage of -18 V. This fit applies to the cylindrical GPM only.  (The Omega GPM  described later in this report has significantly different geometry and would not be expected to follow this curve.)

To account for pressure variations in the course of monitoring the gas purity, it is necessary to measure $P$, compute $J_{fit}$, and to calculate a normalized GPM parameter $S=J/J_{fit}$. An appropriate pressure gauge must be located in close proximity to the GPM so that the pressure corrections can be made.

\section{ The Omega GPM Operation in Oxidation Mode}
\label{sec:OmegaGPM}
\begin{figure}
\centerline{\includegraphics[height=6cm]{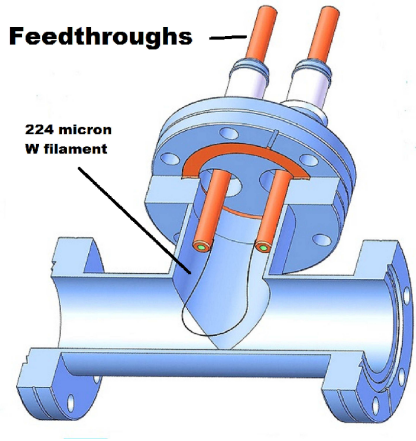}}
\caption{An ``Omega'' style GPM consisting of 2-3/4 conflat vacuum components, a tungsten filament attached to two posts, and a feedthrough on one side. This design has been used to study oxidation effects on the tungsten filament. It can also serve as a purity monitor as described in the text.}
\label{figOmega}
\end{figure}

The Omega GPM has been used to study the effects of impurities on the tungsten filament and as a purity monitor as well. The simplicity of the device shown in Figure \ref{figOmega} is appealing.  All parts can be obtained from catalogs. On the other hand, it lacks the symmetry of the cylindrical device, so comparison of performance cannot be compared quantitatively to analytic calculations.  Its behavior in the presence of impurities must be measured in experiments in the lab.  We have done that for several situations, described next in this section.

\subsection{ Demonstration Experiment}

\begin{figure}
\centerline{\includegraphics[height=6cm]{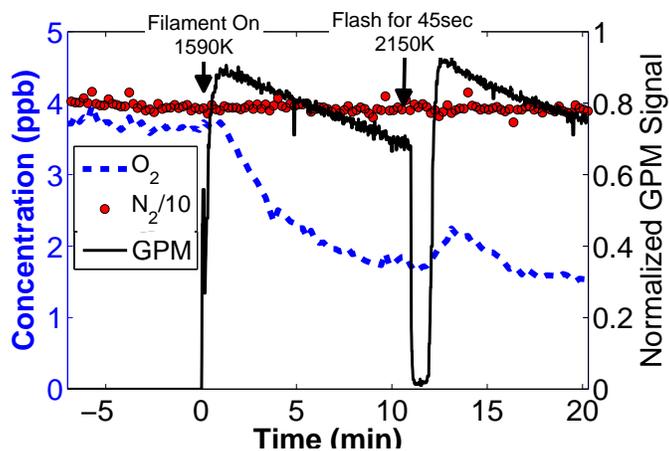}}
\caption{ A demonstration of GPM operation in oxidation mode. In this experiment, xenon gas containing ${\rm O}_2$ and ${\rm N}_2$ flows through the GPM at a flow rate of 0.3 SLPM. Downstream of the GPM we measure the absolute concentration of ${\rm O}_2$ and ${\rm N}_2$ with the coldtrap/RGA technique\cite{bib1}. The GPM filament is turned off until $t=0$, at which time we heat it  to 1600 K. The normalized GPM signal (S) immediately rises to 0.9 and begins to fall linearly. At $t=11$ minutes we degas the filament for 45 seconds, which resets the GPM signal to 0.9. For this particular GPM device and operating conditions, S = 1 corresponds to 400 nA of emission current. See text for further details.}
\label{figDemo}
\end{figure}

We now describe how the GPM can be used to measure gas purity when the oxidation mechanism is dominant. The behavior of the GPM under these conditions is illustrated in Figure \ref{figDemo}. In this experiment, 0.3 SLPM of xenon gas contaminated with ${\rm O}_2$ and ${\rm N}_2$ passes through the Omega-style GPM at 1050 torr, and we measure the absolute concentrations of the two impurity species downstream of the GPM with the coldtrap/RGA technique\cite{bib1}. We find that before the GPM filament is heated the xenon contains 3.8 ppb of ${\rm O}_2$ and 40 ppb of ${\rm N}_2$. At $t=0$, we heat the GPM filament to 1600 K, and the GPM signal rises to S = 0.9 and begins to fall linearly. Since the purity of the xenon entering the device remains constant, we infer that the oxidation mechanism dominates the emission current value for these conditions. This is confirmed by the coldtrap/RGA measurements downstream of the GPM, where we see the ${\rm O}_2$ concentration begin to fall (indicating that oxygen is being removed by the hot filament), while the ${\rm N}_2$ level remains constant.

Since the GPM signal ($S$) is falling, it is necessary to reset the device before the emission current reaches zero. At $t=11$ minutes, we degas the filament by heating it to 2150 K for one minute. During this time, the GPM current drops to zero, which we interpret as evidence that the filament is ejecting impurities from its surface at this elevated temperature, and thus becomes space charge limited. In fact, we see the ${\rm O}_2$ concentration downstream of the GPM increase during and after the degassing process as expected. At $t=12$ minutes, we return the filament to 1600 K, and we find that the GPM signal has returned to 0.9 and is falling linearly as before. We infer from this experiment that the degassing process removes ${\rm O}_2$ from the filament surface, which returns the work function of the tungsten to its original value, after which it begins to oxidize again. Under these conditions, we take the time rate change of the GPM signal ($dS/dt$) as the appropriate figure of merit for the gas purity.

It is important to note that the GPM is actually removing impurities from the gas stream while it measures the gas purity, as shown in Figure \ref{figDemo}. In fact, this cleaning effect was observed as early as 1912 by Langmuir \cite{bib2,bib3}. This local purification effect is most pronounced when the GPM is operated in static xenon, and in this case the xenon within the GPM itself becomes purified as ${\rm O}_2$ is consumed by the tungsten filament and deposited as ${\rm WO}_3$. Also, the buildup of other gases such as ${\rm H}_2$, ${\rm CO}$ and ${\rm CH}_4$ in the device can lead to a false purity measurement if the xenon gas is static. The presence of any of these gases increases the electron velocity causing the emission current to rise \cite{bibd}. When the gas flow rate is a few SLPM or larger, however, the GPM purification effect is small.

\subsection{Calibration of the decay rate vs ${\rm O}_2$ concentration.}

We calibrate the GPM purity measurement in the oxidation mode by measuring the decay rate dS/dt for a variety of gas samples of varying impurity concentrations. As before, the absolute impurity concentration of the xenon gas is determined by the coldtrap/RGA method. This technique was itself calibrated by mixing purified xenon with known amounts of ${\rm O}_2$ and ${\rm N}_2$ \cite{bib1}.

\begin{figure}
\centerline{\includegraphics[height=6cm]{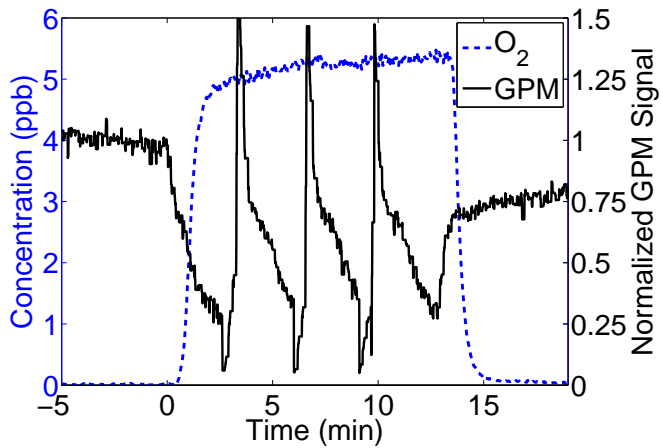}}
\caption{ A calibration of the GPM signal decay rate. In this experiment we measure the absolute concentration of oxygen entering the GPM with the coldtrap/RGA. Prior to $t=0$ the gas entering the GPM has been purified, and the normalized GPM signal is S = 1. At $t=0$ we bypass the purifier, and the GPM signal falls. The RGA measures the ${\rm O}_2$ concentration to be 5.4 ppb. We degas the filament three times at $t=3,\ 6,\ \rm{and\ }9$ minutes and repeat the decay rate measurement. At $t=13$ minutes we begin to purify the incoming gas again, and the GPM signal rises as the ${\rm O}_2$ concentration goes to zero. See text for additional details.}
\label{figOmDKRate}
\end{figure}

An example experiment is shown in Figure \ref{figOmDKRate}. We start with highly purified xenon produced by the purifier, giving a GPM signal of $S=1$ and no signal decay. The pressure in the GPM is maintained at 1050 $\pm$ 30 Torr, and the flow rate is maintained at 4.9 $\pm$ 0.1 SLPM. At $t=0$, we bypass the gas flow around the purifier, and the ${\rm O}_2$ concentration rises to 5.4 ppb. The GPM signal begins to fall. (Note that in these experiments, the RGA/coldtrap is arranged to measure the absolute purity of the gas entering the GPM, rather than the output purity as shown in Figure \ref{figDemo}.) We measure the decay rate $dS/dt$ and then degas the filament. We repeat the process twice, and we find that the decay rate is repeatable to within 5\%.

At $t=13$ minutes, we return the xenon gas system to purify mode, and the ${\rm O}_2$ concentration entering the GPM returns to zero. We then observe that the GPM signal quickly rises to 0.7 and rises slowly thereafter. The rising GPM signal could be interpreted both as evidence for the improving purity, and also as further evidence that the filament is cleaning its own surface by ejecting ${\rm O}_2$, even at the lower temperature of 1600 K. Calculations of ${\rm O}_2$ oxidation and ejection rates are described in Section \ref{sec:OxidationDiscussion}.

\begin{figure}
\centerline{\includegraphics[height=6cm]{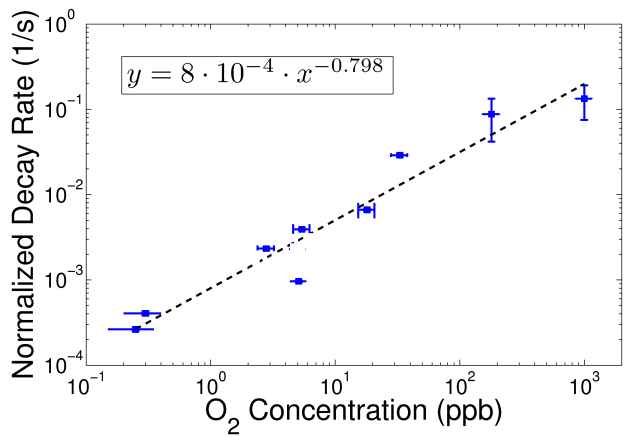}}
\caption{ The normalized GPM signal decay rate (dS/dt) as a function of the absolute ${\rm O}_2$ concentration of the incoming gas. High decay rates correspond to high ${\rm O}_2$ concentrations, as expected.}
\label{figOmDsDt}
\end{figure}

We measured the decay rate $dS/dt$ for a variety of ${\rm O}_2$ impurity concentrations, as shown in Figure \ref{figOmDsDt}. As expected, the signal decay rate is correlated with the true ${\rm O}_2$ concentration over three orders of magnitude, indicating the usefulness of the device as a purity monitor.

\subsection{Reset of the filament by degassing}

In a second experiment, we prepared a xenon gas sample contaminated with 180 ppb of ${\rm O}_2$. Again we start with highly purified xenon produced by the purifier and a GPM signal of $S=1$. At $t=0$ we bypass the purifier, and the GPM signal quickly drops to zero. At $t=2$ minutes we return the system to purify mode, and the GPM signal begins to rise slowly, again indicating that the filament is slowly cleaning its surface by ejecting oxygen. At $t=4$ minutes we degas the filament at 2150 K, after which the signal quickly returns to S = 1, indicating a complete reset of the filament properties. We note that the $S=1$ signal in highly purified xenon is quite reproducible, once the filament has been properly degassed.

\begin{figure}
\centerline{\includegraphics[height=6cm]{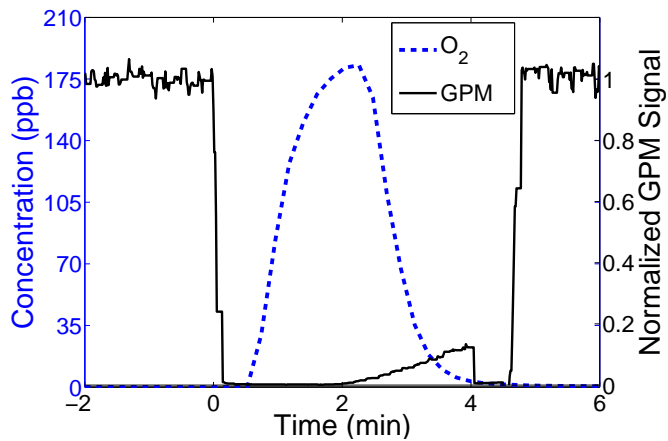}}
\caption{ The effect of degassing on the GPM signal. Prior to $t=0$ the gas entering the GPM is purified, and the normalized GPM signal is S = 1. At $t=0$, we bypass the purifier, and the ${\rm O}_2$ concentration rises to 180 ppb, while the GPM signal quickly drops to zero. At $t=2$ minutes we begin to purifier the gas again, and the ${\rm O}_2$ concentration drops to zero. The GPM signal recovers slowly until $t=4$ minutes, when we degas the filament for 45 seconds by heating it to 2150 K. After degassing the GPM signal returns to $S=1$.}
\label{figDeGas}
\end{figure}

\subsection{Effect of operating at higher filament temperature}
As the degassing experiment illustrates, impurities are being ejected from the filament surface even at the normal operating temperature of 1600 K. Therefore we expect that operating the GPM at elevated filament temperatures would increase the rate of impurity ejection, resulting in a cleaner filament surface and less GPM signal decay. We have confirmed this effect by varying the filament temperature in a series of experiment with identical gas purity. We find that by increasing the filament temperature from 1750 K to 1990 K we reduce the GPM signal decay rate by a factor of eight, as shown in Figure \ref{figDeGas}. This gives further evidence that the GPM emission current is determined primarily by the condition of the filament surface. In light of these results, we can view the degassing process as a limiting case of normal GPM operation, where the GPM filament temperature is high enough that the relevant impurities are no longer able to bond to the filament surface. Measurements of surface interactions between ${\rm O}_2$ and ${\rm N}_2$ indicate that this should occur at 2000-2100 K, as discussed below in Section \ref{sec:GPMResponseToN2}.

\subsection{Absorption of ${\rm O}_2$ on stainless steel}
\begin{figure}
\centerline{\includegraphics[height=6cm]{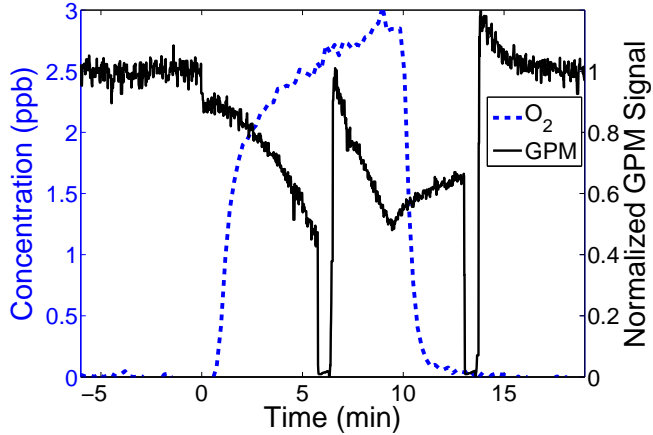}}
\caption{Evidence of absorption of ${\rm O}_2$ on the stainless steel plumbing. Prior to $t=0$ purified gas is entering the GPM, and the normalized GPM signal is $S=1$. At $t=0$ the purifier is bypassed, and the ${\rm O}_2$ concentration rises to 2.8 ppb. The GPM signal drops abruptly to $S=0.9$, and then falls slowly. By $t=5$ minutes, the GPM signal is decaying more quickly. We interpret the slow decay as evidence that ${\rm O}_2$ is being removed from the gas stream by the stainless steel plumbing upstream of the purifier. At t = 6 minutes, 
the filament is degassed, resetting to GPM signal to S = 1, but
the signal continues to fall sharply in the presence of the $O_2$. 
At t = 10 minutes, we remove
the $O_2$ from the input gas using the purifier, and the GPM 
signal begins to rise and level off at S = 0.6. One more degassing
of the filament, at t = 13 minutes, results in the GPM signal returning
to S = 1.}
\label{figSS}
\end{figure}

In a third oxidation-mode experiment with 2.8 ppb of ${\rm O}_2$, we find evidence for ${\rm O}_2$ adsorption on the stainless steel plumbing upstream of the GPM. As shown in Figure \ref{figSS}, we again start with purified xenon and a GPM signal of $S=1$. At $t=0$ we bypass the purifier, and the GPM signal decays as usual. However, the decay rate is slow at first, and becomes faster after a few minutes. A similar effect is seen in xenon gas samples with 0.25 and 0.3 ppb ${\rm O}_2$. At $t=6$ minutes we degas the filament and subsequently reproduce the faster decay rate. We infer from the slow initial decay rate that the ${\rm O}_2$ impurities take several minutes to reach the GPM, although the gas flow rate is large enough to carry them to the GPM very quickly. This indicates that the impurities are temporarily removed from the gas by the interior surface of the stainless steel plumbing between the purifier and the GPM. This surface has been purged by pure xenon gas for a significant period of time before the introduction of the ${\rm O}_2$. Once the surface becomes saturated with the impurity, then the ${\rm O}_2$ begins to reach the downstream GPM.

\subsection{GPM signal overshoot}
As shown in Figures \ref{figOmDKRate} and \ref{figSS}, after degassing the filament the GPM signal sometimes temporarily overshoots the ideal value of S = 1. The origin of this effect is not clear, but we observe this behavior most commonly in highly purified xenon gas. For example, our 180 ppb data, shown in Figure 10, indicates virtually no overshoot at all. It is possible that the GPM filament is putting out small amount of a gas such as ${\rm CO}$ or ${\rm CH}_4$ that is not strongly electronegative, causing the electron drift velocity to increase \cite{bibd}, resulting in elevated currents. In any case, the overshoot is temporary, and it does not affect the subsequent measurement of the decay rate of the signal.

\subsection{Discussion of oxidation}
\label{sec:OxidationDiscussion}
In the absence of space charge, the thermionic current density emitted from a heated tungsten surface is given by Richardson's law: $I \sim T^2e^{-\Phi/kT}$, where $\Phi$ is the work function of the filament surface. The work functions of tungsten and tungsten oxide have been measured to be 4.5 eV and 5.59-5.7 eV respectively \cite{bib4,bib5}. These values imply that a fully oxidized filament will suffer a reduction factor of 1000 to 6000 in thermionic emission compared to a pure tungsten surface.

Degassing the filament at 2150 K for 45 seconds rapidly removes oxygen from the surface and restores the work function of the filament. At 1600 K mainly ${\rm W}_2{\rm O}_6$ is ejected from the filament's oxidized surface at a relatively low rate \cite{bib6,bib7}.  At 2150 K single oxygen along with ${\rm W}{\rm O}_2$ is ejected at 100$\times$ and 10$\times$ the rate of evaporation at 1600 K, respectively \cite{bib6,bib7}.  Once the filament's surface has oxidized its work function will not recover until the oxide layer has been desorbed.

Some basic considerations allow us to infer the rates of oxygen capture and ejection from the filament surface in 1050 torr of xenon at 1600 K. In Figure \ref{figDemo}, xenon containing 3.8 ppb of ${\rm O}_2$ flows at 0.3 SLPM across the GPM filament. This corresponds to $1.25 \times 10^{14}\  {\rm O}_2$ molecules per minute. When the filament is heated, the ${\rm O}_2$ concentration at the GPM outlet is reduced by 58\%, indicating an ${\rm O}_2$ capture rate of $7.3 \times 10^{13}$ molecules per minute. On the other hand, we estimate that our filament has about $5  \times 10^{14}$ tungsten atoms on its surface, and the observed emission current decay rate indicates that the filament surface will fully oxidize in 50 minutes. This corresponds to a net surface coverage rate of $6.7 \times 10^{12}$ molecules per minute. Since the net coverage rate is observed to be much smaller than the ${\rm O}_2$ capture rate, we infer that the difference is due to ejection of tungsten oxide (predominantly ${\rm W}{\rm O}_3$ and ${\rm W}_2{\rm O}_6$ \cite{bib6}) from the filament surface at a rate of $6.6 \times 10^{13}$ molecules per minute in 1050 torr of xenon. Other studies of tungsten filaments in vacuum find ejection rates which are four times larger than this\cite{bib6}.

\subsection{Formation of tungsten-oxide powder in the GPM}
A major concern with burning a tungsten filament in impure gas is the formation of byproducts such as tungsten oxide. This substance is continually emitted from the GPM filament during operations, and it could travel with the gas stream and contaminate the rest of the apparatus unless it is removed with a particulate filter. Secondly, if the oxygen concentration is as high as $10^{-6}$, then the resulting tungsten oxide can quickly cover the conflat flange and short the GPM filament leads.

The color of the tungsten oxide is indicative of the concentration of the ${\rm O}_2$ in which it was formed. For 10 ppb, the color is grey-white, for 100 ppb it is blue-white, and for 1 ppm it is black. At $10^{-4}$ the color is yellow, and the filament will burn out within a few minutes. The blue-white powder found inside the GPM was analyzed using X ray crystallography, and its composition was found to be 90\% ${\rm WO}_3$, 8\% ${\rm WO}_2$ and 2\% ${\rm W}_{18}{\rm O}_{49}$. The non-stoichiometric defects are responsible for the color of the tungsten-oxide crystal. These observations are consistent with results reported by other groups while growing thin tungsten oxide films in oxygenated environments using argon sputtering \cite{bib13,bib14,bib15}.

During the study, both the blue and the black tungsten oxide caused electrical shorts in the GPM. The white powder formed at low ${\rm O}_2$ concentrations is amorphous and might not be conductive \cite{bib15}. In order to stop the tungsten oxide powder from spreading through the system and potentially contaminating the detector, particulate filters (Mott filters) must be used at the output of the device. Without Mott filters we have observed the tungsten oxide powder traverse several meters in our plumbing. The diameter of the blue tungsten oxide particulates is as small as 10 nm \cite{bib14}. The majority of the tungsten oxide particles observed in the plumbing after oxidization tests are on the order of several microns. [section 4.8]

\begin{figure}
\centerline{\includegraphics[height=6cm]{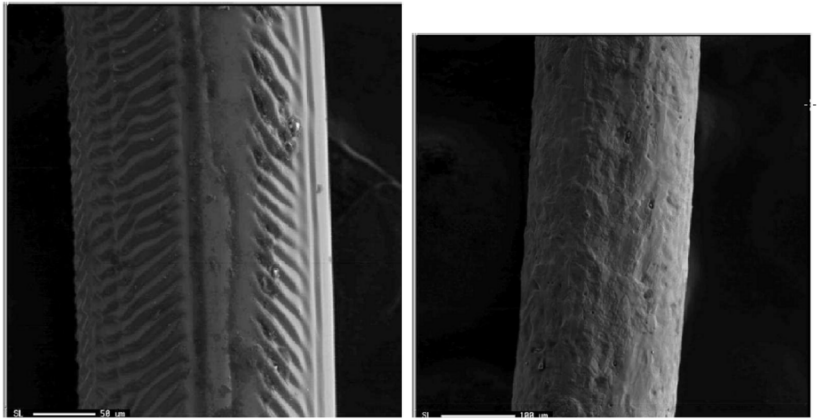}}
\caption{Left: Filament after burning in static xenon initially containing 1ppm ${\rm O}_2$ for 150 hours. The evaporation pattern is visible. Right: Filament after being used for 2 months with purified xenon (0.1ppb) ${\rm O}_2$. After burning in xenon the filament is more brittle. The tungsten re-crystallizes in a granular structure with the grains ranging in radius from 4 to 20 microns in diameter.}
\label{figFilBurn}
\end{figure}

\begin{figure}
\centerline{\includegraphics[height=6cm]{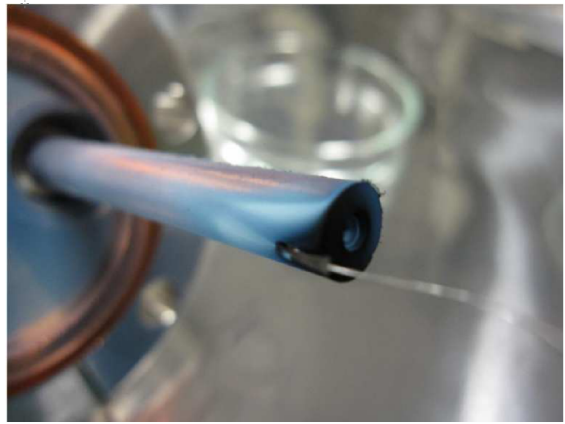}}
\caption{Tungsten oxide formed in 1350 torr of static xenon in a 0.58 L volume with 100 ppb oxygen}
\label{FigWO2}
\end{figure}

Figure \ref{figFilBurn} show SEM images of the filament after burning in 1 ppm ${\rm O}_2$ and one that was used in purified xenon. The device was used with pressures of 1350 Torr. Figure \ref{FigWO2} shows electrodes coated with tungsten oxide.

\section{GPM response to ${\rm N}_2$}
\label{sec:GPMResponseToN2}
The xenon gas samples used in our GPM calibration experiments contained some ${\rm N}_2$ contamination, so we investigated the effect of nitrogen on the performance and behavior of the device. First, we note that our demonstration experiment, shown in Figure \ref{figDemo}, indicates that ${\rm N}_2$ is not removed from the gas when the GPM filament is heated. Secondly, we prepared a sample of xenon containing significant ${\rm N}_2$ and less than 0.1 ppb of ${\rm O}_2$.
 
The GPM signal under these conditions is $S=1$, and does not change when switching from bypass mode (${\rm N}_2$ contamination of 20 ppb) to purify mode (${\rm N}_2 < 1$ ppb). We conclude that the presence of nitrogen has little effect on the GPM signal. This result is not surprising, because at temperatures above 1000K ${\rm N}_2$ adsorption onto tungsten ceases and ${\rm N}_2$ does not bond to a tungsten surface \cite{bib8}.

The GPM results presented above indicate that the device is not sensitive to the presence of ${\rm N}_2$ in the xenon gas. This shortcoming is acceptable in most situations because ${\rm N}_2$ presents  less of a problem to TPC operation than does ${\rm O}_2$ due to its much smaller electron capture cross section\footnote{Nitrogen electron capture rates $\sim$ 800 times smaller than oxygen have been reported\cite{bibe}.}.
Nevertheless, xenon which is contaminated with ${\rm O}_2$ usually contains even higher concentrations of ${\rm N}_2$, due to its larger presence in air. Therefore it would be useful to have a complementary technique for observing ${\rm N}_2$ as a means of inferring the presence of ${\rm O}_2$. Secondly, noble gas purifiers such as the zirconium getters used by EXO have ten times less capacity for absorbing ${\rm N}_2$ than ${\rm O}_2$. Therefore the presence of significant ${\rm N}_2$ in the xenon gas system could be an indication that the purifier is more than 10\% exhausted.

${\rm N}_2$ impurities cannot be detected with the GPM at 1600 K because nitrogen desorbs rapidly from the filament between 1000-2300 K \cite{bib8,bib9}. In principle, ${\rm N}_2$ could be detected by increasing the filament temperature to 2300 K, where another phase of tungsten nitride begins to form\cite{bib8}. Under these conditions, we would expect the GPM signal to decay as ${\rm N}_2$ is adsorbed on he filament. In practice, however, we find that our GPM emission current becomes space charge limited under these conditions, and is then independent of the work function.

\begin{figure}
\centerline{\includegraphics[height=6cm]{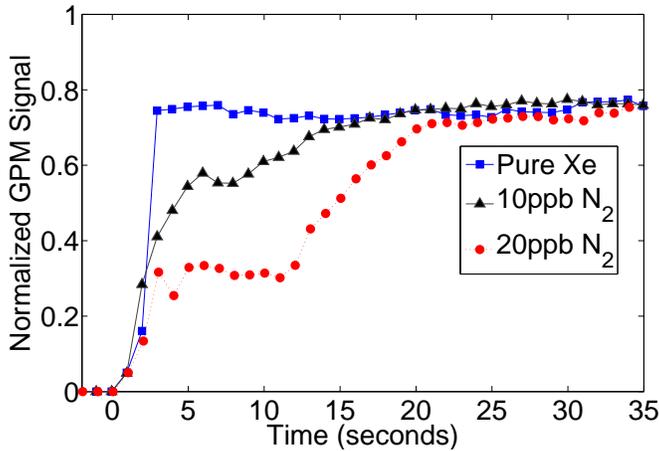}}
\caption{ An experiment to observe adsorption of $N_2$ and other gases on the GPM filament at room temperature. The GPM filament is left at room temperature immersed in the gas of interest for 12 hours. At t = 0 we turn on the filament, and the emission current rises as $N_2$ is ejected from the surface and the work function decreases. The rise time could be used to infer the quantity of these impurities in the gas. See text for further details.}
\label{FigN2adsorp}
\end{figure}

At room temperature, however, a pure tungsten filament will adsorb enough ${\rm N}_2$ to cover a surface layer\cite{bib8}. Tungsten nitride is expected to cause the emission current to suffer a factor of 50 reduction at 1600 K due to the increased surface work function \cite{bib10}. ${\rm H}_2{\rm O}$, ${\rm H}_2$, and other hydrocarbon impurities are also expected to adsorb at room temperature \cite{bib11,bib12}. Therefore we performed a series of experiments to attempt to observe ${\rm N}_2$ absorption on the surface of the GPM filament at room temperature. In these experiments we left the GPM filament off while immersed in a static xenon gas sample overnight. When using a xenon gas sample containing $ 1 \times 10^{-9}$ g/g of ${\rm N}_2$, the GPM signal rises to $S=0.8$ within 3 seconds after heating to 1600 K. In other xenon samples containing $10\times 10^{-9}$ g/g and $20 \times 10^{-9}$ g/g of ${\rm N}_2$, however, the rise time of the signal is 16 seconds and 22 seconds, respectively, as shown in Figure \ref{FigN2adsorp}. The longer rise times observed in these datasets are consistent with the presence of a layer of ${\rm N}_2$ adsorbed on the filament surface, which would temporarily reduce the tungsten work function until the surface is cleaned of this impurity. It should be noted, however, the observed effect could be due to some other impurity such as a hydrocarbon, which would also desorb rapidly at 1600 K. Nevertheless, this effect could be useful for detecting impurity species which are less electronegative than ${\rm O}_2$.

\begin{figure} [t]
\centerline{\includegraphics[height=6cm]{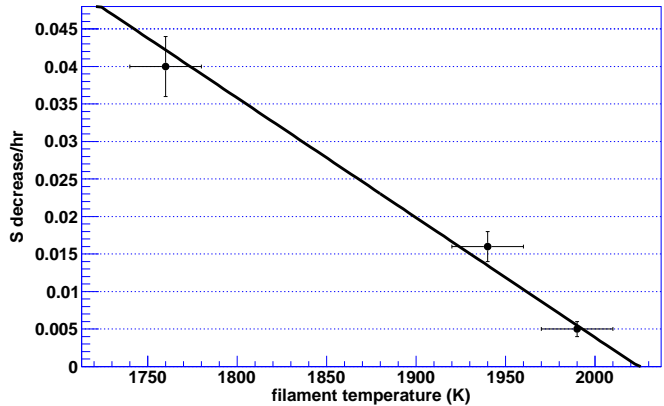}}
\caption{ Decay rates of the cylindrical GPM current versus filament temperature from Nitrogen impurities. The solid line is a linear fit to the data.}
\label{figGPMN2DK}
\end{figure}

For the cylindrical GPM, it is difficult to observe any effect of nitrogen contamination up to the $\sim$ 1 ppm level. However, for nitrogen contaminated xenon, while GPM current is constant in time for the default 4.0 A setting, a decay with time of the GPM response is observed for lower filament currents. A test was performed in which a fixed dosage of 1 ppm nitrogen was added to the gas flow, and the GPM response decay time was observed as a function of the GPM filament current\footnote{Note that in these experiments, the nitrogen dosed into the system was controlled at the $<$10\% level for all data points, but the nitrogen content of the gas flow was not independently measured.}.
The filament current setting was varied between 3.0 A and 4.0 A. The GPM current decay rate was observed to fall to zero as the 4.0 A setting was approached.  We have used the known tungsten thermal resistivity coefficients, and the observed power supply voltage at each filament current setting, to deduce the filament temperature at each filament current setting. Figure \ref{figGPMN2DK} shows the decay time data as a function of filament temperature.  The GPM current decay rate falls to zero at a filament temperature of about 2000 $\pm$ 50 K, corresponding to a filament current of $\sim$ 3.6 A. This behavior suggests that at a characteristic temperature, nitrogen is completely desorbed from the tungsten filament surface. The temperature observed here corresponds closely to values for nitrogen desorption from tungsten in the literature \cite{bib16}.

\section{Tests with Argon}

\begin{figure}
\centerline{\includegraphics[height=6cm]{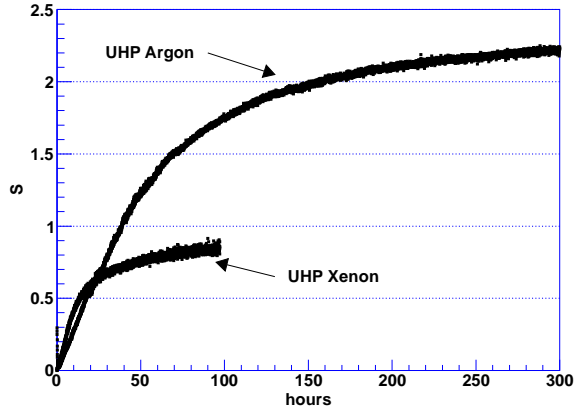}}
\caption{Purification of UHP grade argon compared to UHP grade xenon in the closed-loop test system. The GPM currents have been normalized in both cases to the value for pure xenon. The higher current with argon is expected because of the higher electron drift velocity than in xenon.}
\label{figArgXen}
\end{figure}

It is reasonable to extend these principles and studies to other noble gases. We have only done this, briefly, with argon. Neon and helium have not been studied. The test with argon was a simple one, namely to purify a sample of ultra-high-purity (UHP) argon in our system, circulating it through the  purifier. UHP grade argon from a commercial supplier comes with specifications on certain impurity levels.  Our sample had the following limits on impurities:
$<$ 1 ppm ${\rm O}_2$; $<$ 0.5 ppm ${\rm H }_2{\rm O}$; $<$ 1 ppm CO; $<$ 1 ppm C${\rm O}_2$; and $<$ 5 ppm ${\rm N}_2$.
For comparison we have a comparable test on UHP grade xenon.  The specifications on purity for the xenon sample are essentially the same. Figure \ref{figArgXen} shows the cylindrical GPM response for the duration of the test, with the UHP xenon data superimposed.  Note that the GPM current in both cases has been normalized to the value for pure xenon. This illustrates the fact that the GPM current in argon is approximately a factor of two larger than for xenon under the same conditions. This is due to the higher electron drift velocity in argon.

One more point should be emphasized about these data.  The flow of gas was such that several volume changes per hour occurred.  We know (and have observed) that the purifiers remove almost all impurities on a single pass, at these rates.  Why then does it take so long to achieve high purity?  The reason is that the impurities have migrated to the surfaces in the system, remain there until they slowly are removed into the flowing gas stream. In addition, if readily permeable substances such as plastics are in the system, impurities are absorbed into the bulk material. The time constants are set by the rate of desorption from the surfaces and outgassing from materials, not by the flow rate of the gas.

\section{GPM Operation with EXO-200}
\begin{figure} [t]
\centerline{\includegraphics[height=6cm]{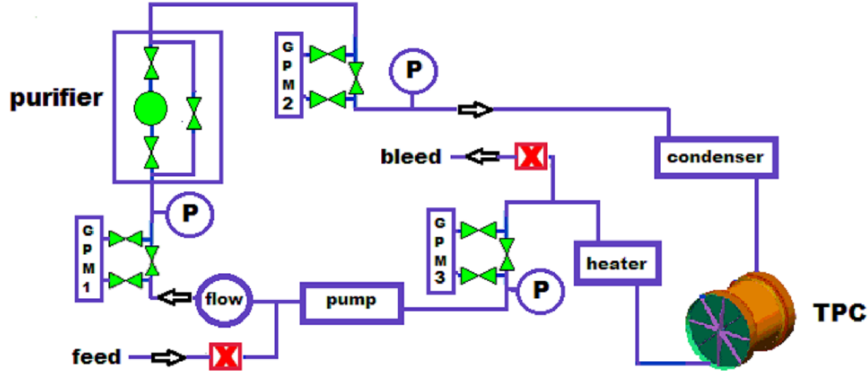}}
\caption{A simplified diagram of the gas loop for EXO-200.  Three GPMs monitor the purity of the gas as it circulates. In close proximity of each is a pressure gauge.  The manual valves associated with the GPMs and the purifier allow for bypassing these components when desired.  Two computer driven valves on the inlet (feed) and outlet (bleed) allow for control of the system pressure to stringent tolerances.}
\label{figSimpEXO}
\end{figure}

Three cylindrical GPMs of the design shown  in Figure \ref{figCyl} have been in use at EXO-200.  Figure \ref{figSimpEXO} is a simplified diagram of the gas loop of the system. The pump creates a flow up to 20 SLPM.  This flow passes first through a flow meter, then through GPM-1 before the purifier\footnote{SAES Pure Gas, Inc. Model PS4-MT3-R; two in parallel}. GPM-2 is located immediately after the purifier and monitors the health of the purifier. GPM-3 is located in the output leg of the TPC, and monitors the purity of the xenon after passing through the TPC. The condenser and heater are used when the TPC has liquid xenon in it; otherwise they are passive when the TPC has only xenon gas in it. The system has operated reliably in both states during commissioning of EXO-200.

\subsection{ Practical considerations for using a GPM }
There are practical considerations involved in using a GPM to monitor the purity of the gas. As shown in Figure \ref{figScopes}, the response time to electronegative impurities is several seconds. The rate at which one can monitor  a gas stream can be as short as a few seconds, or as infrequently as conditions require.  During tests in the lab we typically sample the gas purity once every 5 minutes. For EXO-200 at WIPP, the sample rate has varied from once every 20 minutes to once every few hours, depending on circumstances at the time.

The GPM is very sensitive to impurity concentrations and the act of sampling the purity can disturb the reading. Impurities reside on all surfaces, including the GPM walls. When the filament is on, $\sim$40 W of power is transferred to the gas flow by convection and conduction, while at the same time the local surfaces are radiatively heated. The rising temperature causes impurities to evolve into the local gas, distorting the readings.  To mitigate this problem, we keep the sample duration as short as possible: around 10-15 seconds. Also we  have enclosed the GPMs in heater jackets  which maintain a steady GPM body temperature of $\sim$ 100 deg C, to reduce these transient heating effects.

As discussed in Section \ref{sec:OmegaGPM}, oxidation or similar effects can occur at the surface of the filament. This can result in filament burnout and the need for replacement. We have replaced the filaments on several occasions in the lab and at WIPP. To facilitate the filament replacement, the GPMs have 3 valves attached,  one for the input side, one for the output side, and a bypass valve. In addition, the bottom  ``tee''  shown in Figure \ref{figCyl} is actually a ``cross'' in the EXO-200 GPMs at WIPP. The extra port provided by the cross has a copper pinch off tube mounted on it.  It is the port used for pumping out the GPM for a vacuum bakeout following a filament replacement.

To minimize the damage to the filaments, The control system stretches out the time between samples when the normalized signal $S$, falls below 0.1.

\subsection {Dependence on the flow rates}
During operations of the EXO-200 cylindrical GPMs in space-charge limited mode at WIPP, it was observed that high flow conditions affected the observed GPM currents.  Initially, the observations showed reduced GPM current in purified xenon as the gas flow rate was increased.  It was subsequently also seen that high flow conditions led to elevated GPM currents in impure xenon gas.

\begin{figure}[t]
\centerline{\includegraphics[height=6cm]{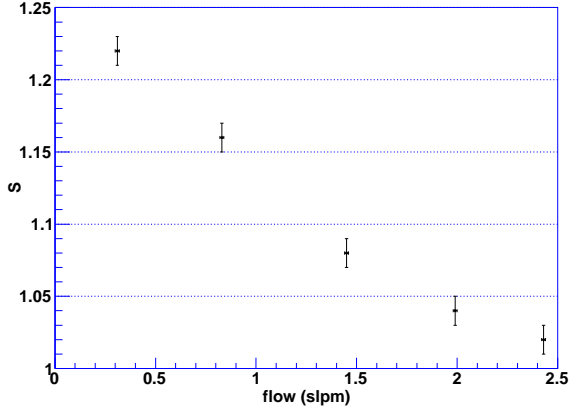}}
\caption{The normalized cylindrical GPM current ($S$) is plotted as a function of flow rate in purified xenon. Our normal operating point in the lab is $\sim$1.5 SLPM. At WIPP, the flow rate is higher.}
\label{Figflowdep}
\end{figure}

We studied the effect of gas flow rates on the cylindrical GPM performance in our lab setup. Figure \ref {Figflowdep} shows how normalized GPM current is affected by the flow rates up to $\sim$ 2.4 SLPM in purified xenon. The effect at the flow rates we achieve in the lab is generally smaller than observed at WIPP (our EXO-200 pump can deliver up to 20 SLPM ).
 
We have considered three possible flow-dependent mechanisms :
\begin{enumerate}
\item Flow- and impedance-dependent pressure drops between the GPM itself and the nearby pressure transducer used to make the pressure dependent normalization of the GPM current.  Depending on the location of the pressure transducer, this can cause a positive or negative shift in the normalized GPM current.
\label{enum:PressureDrops}
\item Flow-dependent convective cooling of the GPM filament.  This effect might change electron emission or impurity desorption from the filament.
\label{enum:Cooling}
\item Flow-dependent displacement or removal of the slow-moving negatively charged impurity ions. Ion velocities in the xenon are nominally around 10 cm/s for typical ion mobilities $\sim$ 1 V ${\rm cm}^{-2}{\rm sec}^{-1}$; this compares with $\sim$ 15 cm/s flow rates at 10 SLPM in the GPM. This effect is expected to increase the observed GPM current.
\label{enum:IonRemoval}
\end{enumerate}

\begin{figure}[t]
\centerline{\includegraphics[height=6cm]{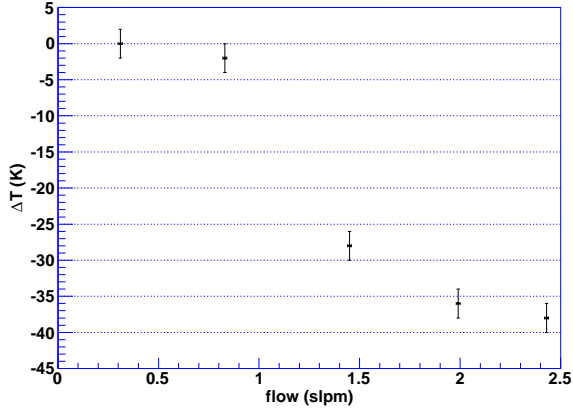}}
\caption{Filament cooling as a function of flow rate in purified xenon. Temperature is determined from the observed filament resistance.}
\label{Figfiltemp}
\end{figure}

Effect \ref{enum:PressureDrops} must be carefully evaluated in any GPM setup, but as it turns out neither at EXO-200 nor in our lab is this effect sufficiently large to explain the observed GPM shifts, typically by at least one order of magnitude.  Effect \ref{enum:Cooling} is observed (Figure \ref{Figfiltemp}), where we have deduced the filament temperature from the observed filament resistance and the known thermal resistivity curve for tungsten (2150K is taken the absolute reference point at lowest flow). We are unable to evaluate at this time if these small temperature reductions are causative. The final Effect \ref{enum:IonRemoval}, whereby impurity ions are swept out of the GPM gas, has not yet been studied in the lab.

Because of the flow effects on the GPM currents, the flow rates should be kept low. In our lab, the flow rate is typically 1.5 SLPM. We also avoid using the GPMs in static xenon because the build-up of gases such as H$_2$, CH$_4$, and CO can lead to false purity measurements. At WIPP, since our pump is capable of higher speeds, we usually run the GPMs with bypass valves in parallel open to reduce the flow through the GPMs themselves, and generally avoid maximum flow operation.

\subsection{ Examples }
\begin{figure}
\centerline{\includegraphics[height=6cm]{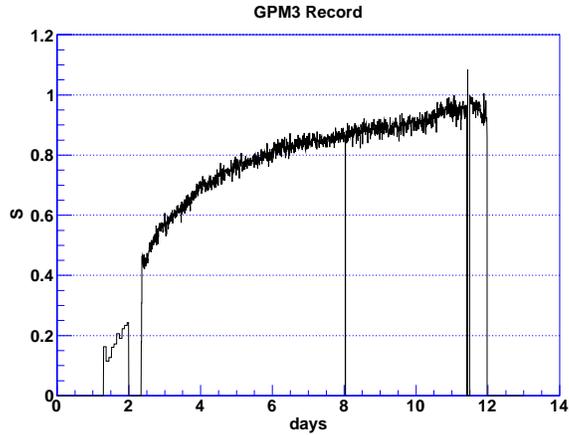}}
\caption{A 11-day period of monitoring the xenon purity in the EXO-200 system at WIPP. Xenon gas flows continuously at $\sim$10 SLPM for the purpose of cleaning the TPC. The GPM at the output of the TPC initially reports a low value for $S$, around 0.1, that rises to $\sim$ 0.95. }
\label{FigatWIPP1}
\end{figure}

We illustrate the use of the GPMs during commissioning of EXO-200 at WIPP with the example of a 11 day period of flowing xenon gas  through the TPC for the purpose of cleaning it, as the TPC contains significant metallic surface area as well as outgassing plastic parts (teflon, kapton, and acrylic). The gas-phase loop of the EXO-200 system shown in Figure \ref{figSimpEXO} consists of (i) a circulation pump which at this time was creating a flow of $\sim$ 10 SLPM , followed by (ii) GPM-1, followed in turn by (iii) two purifiers in parallel, then (iv) GPM-2, (v) the TPC, and (vi) GPM-3.  After GPM-3 the gas returns to the input side of the pump. The xenon was sampled continuously by  GPM-3 situated in the output of the TPC. Figure \ref{FigatWIPP1} shows its response. Initially, the monitor current is suppressed, indicating rather poor purity. The signal grows steadily over the next 11 days, reaching a good value of $S$ $\sim$ 0.9. The GPM cannot distinguish the species of impurity, nor can we determine its concentration. Subsequent analysis of gas samples, however, indicated presence of trace amounts of  ${\rm H}_2{\rm O}$ and isopropanol (which was used for cleaning parts).

\begin{figure}
\centerline{\includegraphics[height=6cm]{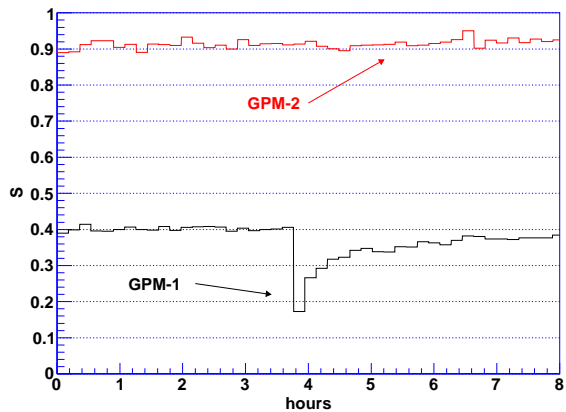}}
\caption{A 4-hour period of monitoring the xenon purity in the EXO-200 system at WIPP. The injection of a small amount of gas from supply bottles indicates some impurity accompanies the gas, as shown by the response of GPM-1. The gas flows through the purifiers and through GPM-2, which shows no response, indicating that the  purifier is operating effectively. }
\label{FigatWIPP2}
\end{figure}

A second example of the use of the GPMs during commissioning of EXO-200 at WIPP comes from a 4-hour period when xenon gas was circulating in the system and the TPC had no liquid in it. Figure \ref{FigatWIPP2} captures a typical period of monitoring, chosen here because a ``feed'' event occurs during this period. A feed cycle is simply an injection of a small amount of xenon from storage bottles to maintain the TPC pressure within a tightly defined narrow band. The gas from the storage bottles, however, also contains some small impurity component.

The feed of xenon is introduced into the loop after the pump and before GPM1. Figure \ref{FigatWIPP2} shows this occurring a little before 4 hours. The responses of  two GPMs are shown. GPM-1 shows a drop in $S$, the normalized signal, due to impurities in the feed gas.  Although the feed event lasts only seconds, the GPM-1 takes $\sim$ 2 hours to recover.  The slow response is due to impurities accumulating on the walls and surfaces in the vicinity of GPM1. The flowing xenon slowly picks up the impurities and carries them through the purifiers. GPM2, situated after the purifiers, however shows no change. The purifiers have removed all of the impurities from the xenon passing through. From these data we learn; (i) that the feed stock from the storage bottles is impure; (ii) the purifiers are functioning well.

These examples show the use of the GPMs at WIPP.  They serve well as online monitors of the health of the gas loop in the EXO-200 system.

\section{Conclusion}
We have constructed and characterized two types of gas purity monitors which are capable of detecting common impurities in a xenon gas stream with high sensitivity. We have deployed three of these devices in the xenon gas handling system of the EXO-200 experiment, and they have provided gas purity information during the commissioning phase
of the experiment. Most importantly, the GPMs can be operated continuously without disturbing detector operations, giving instant and remotely accessible information about the xenon gas purity at all times.

\label{}
\vskip 0.25 in
{\noindent \bf Acknowledgments}
EXO is supported by the Department of Energy and the National Science Foundation in the United States, NSERC in Canada, FNS in Switzerland and the Ministry for Science and Education of the Russian Federation in Russia. The SLAC National Accelerator Laboratory operates under Department of Energy contract number DE-AC02-76SF00515.

\appendix
\section {A model relating the space charge limited current and the impurity concentration in cylindrical geometry}

In Section \ref{sec:SuppressionByNegIons} we wrote that, under reasonable assumptions, the density of the {\it neutral} impurity component, $M^{\circ}$, is related to the normalized GPM anode current $S$ by
\[M^{\circ} =k {1-S \over S}\]
where $S = J/J_0$  and $J_0$ is the current for pure xenon, and
\[k \approx { 2 e v_{\circ} \mu V  \over \alpha b^2}\]
where $e$ is the charge of the electron, and $\alpha$ is related to the probability for an electron attaching to a $M^{\circ}$ molecule, converting to an $M^-$ ion.
 
To derive this relationship requires cylindrical symmetry, so would not necessarily apply to the Omega GPM design. We make some very reasonable simplifying assumptions.  They are:
\begin{enumerate}[i)]
\item Steady state current flow has been established.  This occurs in the lab tests several seconds {\it after} the filament current has been turned on.

\item The filament temperature is high enough so that the electron current is not emission limited.

\item The ion current can be neglected. The ratio $J_{ion}/J_{electron}$ from (\ref{eq:SpaceChargeLimitedCurrent}) and (\ref{eq:IonCurrent}) in Section \ref{sec:SpaceChargeLimitedCurrents} is $\approx \mu V/v_e b \approx v_{i}/v_{electron}$.  For $\mu \approx 1 {\rm V}\ {\rm cm}^{-2}\ {\rm}s^{-1}$, and a bias voltage of 18 V, this ratio is $\approx 10^{-4}$.

\item The electron velocity is constant. This is a good assumption in this situation because the electric field is nearly constant for $r \gg a$ at a value $V/b$.

\item Negative ions are formed by capturing electrons from the current flow. We assume this process is proportional to the density of neutral impurities $M^{\circ}$ and the density $n_e(r)$ of electrons. This assumption can be expressed as
\[M^-(r) = \alpha\  n_e\  M^{\circ}\qquad {\rm per\  second\  per\  unit\  of\  volume}\]
where $\alpha$ is a ``probability coefficient'',  $M^-$ is the rate of formation of negative ions per unit volume, $n_e$ is the electron density,  and $M^{\circ}$ is the impurity density. The parameter $\alpha$ is a constant in the analysis here, but in the lab its value likely will depend on the species of impurity, the pressure, the gas temperature, and possibly other factors. These global parameters must be held fixed for the relationship between $M^{\circ}$ and $S$ to hold.

\item Steady state ion distribution is established when the number of ions flowing to the anode equals the number of ions formed in the volume. This can be expressed in a mass flow equation. Define a vector for the flow of ions:
\[\overrightarrow {\bf M} = M^- v_i {\bf e_r}\]
where $M^-$ is a (scalar) density of negative ions moving with radial velocity $v_i$ in the direction  ${\bf e_r}$. Steady state conditions occur when:
\[\nabla \cdot \overrightarrow{\bf M} = \alpha \ n_e \  M^{\circ}.\]
The density of electrons is given by
\[n_e = {J  \over 2 \pi r v_{\circ} e}\]
so integration gives $ M^-  = {\alpha J M^{\circ}/ \  2 \pi e  v_{\circ} v_{i}}= {\alpha J M^{\circ}b /\  2 \pi e v_{\circ} \mu V } $, a constant independent of $r$. It scales linearly in $\alpha$ and  $M^{\circ}$ and inversely with $v_{i}$.
\end{enumerate}

 We can now solve for the current-voltage relation using the same space charge analysis we used in deriving (\ref{eq:SpaceChargeLimitedCurrent}) and (\ref{eq:IonCurrent}) in Section \ref{sec:SpaceChargeLimitedCurrents},
\[\nabla \cdot {\bf E} = {\rho_e \over \epsilon_{\circ}} + {M^- \over \epsilon_{\circ}}\]

With the boundary conditions  $\Phi = 0$ and $d\Phi/dr = 0$ at $r = a$ and $\Phi = V$ at $r=b$, we get
\begin{eqnarray*}
V &=& {J (b-a-a\ ln\  b/a) \over 2\pi\epsilon_{\circ} v_{\circ}} + {M^- ((b^2-a^2)/2-a^2\ ln(b/a) \over 2\epsilon_{\circ}}\\
&= &c_1 J + c_2 M^{\circ} J \hskip 2.9 in
\end{eqnarray*}
where $ c_1 \approx   {b / \  2\pi \epsilon_{\circ} v_{\circ}} $ and $ c_2  \approx {\alpha b^3 / \  4 \pi \epsilon_{\circ} e v_{\circ} \mu V}. $ Inverting, we have
\[J = {V \over c_1 + c_2 M^{\circ}}\  {\rm and}\  J_{\circ}\  (J \ {\rm  at}\  M^{\circ}=0)= {V \over c_1}.\]

Defining the normalized GPM current as $S = J / J_{\circ}$ , then
\[S = {1 \over 1+ c_2/c_1\ M^{\circ}}\]
or
\begin{equation}
M^{\circ} = {c_1 \over c_2} {1-S \over S}
\end{equation}
with
\begin{equation}
c_1/c_2 \approx { 2 e v_{\circ} \mu V  \over \alpha  b^2}
\end{equation}
\vskip 0.1 in





\bibliographystyle{model1-num-names}
\bibliography{gpm-bib}


\end{document}